\documentclass[prd,aps,floats,preprint,preprintnumbers]{revtex4}

\usepackage{amsmath,amssymb}
\usepackage{graphicx}
\usepackage{psfrag}

\def\be{\begin{equation}}
\def\ee{\end{equation}}
\def\bea{\begin{eqnarray}}
\def\eea{\end{eqnarray}}
\newcommand{\tmH}{\mathcal{H}}
\def\GB{R_{\text{\sc GB}}^2}

\def\bGB{\bar R_{\text{\sc GB}}^2}
\def\bGBo{\bar R_{\text{\sc GB}0}^2}
\def\N{\nabla}
\def\a{\alpha}
\def\b{\beta}
\def\s{\sigma}

\begin{document}

\setlength{\unitlength}{1mm}
\title{Observational Bounds on Modified Gravity Models}

\author{Antonio De Felice$^1$, Pia Mukherjee$^1$, Yun Wang$^2$}
% Just alphabetical for now

\affiliation{$^1$Department of Physics \&\ Astronomy, University of Sussex,
Brighton BN1 9QH, United Kingdom.\\
$^2$Homer L. Dodge Department of Physics \&\ Astronomy, University of Oklahoma,
Norman, OK 73019, USA.}

\begin{abstract}
  Modified gravity provides a possible explanation for the currently observed
  cosmic accelaration.  In this paper, we study general classes
  of modified gravity models.  The 
Einstein-Hilbert action is modified by using general functions
  of the Ricci and the Gauss-Bonnet scalars, both in the metric and in
  the Palatini formalisms.  We do not use an explicit form for the
  functions, but a general form with a valid Taylor expansion up to
  second order about redshift zero in the Riemann-scalars.  The
  coefficients of this expansion are then reconstructed via the cosmic
  expansion history measured using current cosmological observations.  These
  are the quantities of interest for theoretical considerations
  relating to ghosts and instabilities. We find that current data
  provide interesting constraints on the coefficients. The
  next-generation dark energy surveys should shrink the allowed
  parameter space for modifed gravity models quite dramatically.
\end{abstract}

\maketitle

\section{Introduction}
The discovery of the current accelerated expansion of the Universe 
\cite{Riess98,Perl99} has 
generated a lot of excitement in the last few years. It turns out that 
this unexpected behaviour can be modeled in many ways, most simply 
through a cosmological constant. On the other hand it has been argued that, 
since this behaviour appears on large scales, the acceleration of the 
universe is due to some modification of gravity on such scales.

There are quite a few models which can describe such modifications of gravity, such as
scalar-tensor theories \cite{Dicke:1961gz}-\cite{DeFelice:2005bx}, the so called $f (R)$ theories \cite{Carroll:2003wy}-\cite{Song:2006ej} (which are a subset of
scalar tensor theories), brane world models of which Randall-Sundrum models \cite{Randall:1999ee}
and DGP models \cite{Dvali:2000hr,Deffayet:2001uk,Deffayet:2001pu,Koyama:2005tx} are special cases, and some more complicated function of
curvature invariants, such as $f (R, \GB)$ \cite{Carroll:2004de,Mena:2005ta,Navarro:2005gh,DeFelice:2006pg,Calcagni:2006ye} where
$\GB$ represents the Gauss-Bonnet combination defined as $R^2 - 4
R_{\alpha \beta} R^{\alpha \beta} + R_{\alpha \beta \gamma \delta} R^{\alpha
\beta \gamma \delta}$.

Modifying gravity in a consistent way is not an easy task, hence many of
these models face quite stringent theoretical bounds which can reduce the
parameter space of these theories. In the same way one can try to use 
data to restrict further the parameter space and possibly rule out some of
these models.

In this paper we study $f(R)$ and $f(R,\GB)$ models, most generally, without specifying
any explicit form for these functions. In the $f(R)$ case, we consider both the metric based, and the
Palatini formalisms. 
We use the recent expansion history of the universe, reconstructed
allowing the Hubble parameter to be a free function in redshift bins, using type Ia supernovae (SNe Ia) data together with 
relevant contraints from the cosmic microwave background (CMB) and galaxy surveys, and invert this
to obtain constraints on the parameters of the $f(R)$ models.  This perspective has not
been adopted before. Although it is similar in nature to the reconstruction of the 
quintessence potential \cite{Sahlen:2006dn}, here the scalar degrees of freedom come entirely from the gravity
sector.

Much work has been dedicated to the subject of constraining $f(R)$ theories using solar system 
measurements \cite{Hu:2007nk}-\cite{Allemandi:2005tg}. We will not study this here for several reasons. First, as believed by 
some (e.g.\ \cite{Mota:2003tm}), the cosmological parameters may be different from those measured locally, 
because as the background changes from a locally spherically symmetric metric to a 
homogeneous and isotropic one, the behaviour of quantities such as the coefficients of a Taylor-expansion of $f(R)$ might be 
quite different. Second, for this theory locally (at the solar system scale) the weak field approximation may not hold, 
so that it cannot be matched to a perturbation of GR; i.e., the usual 
constraints cannot be trivially applied because these are found by assuming the metric 
to be a perturbation about the GR-Schwarzschild solution \cite{Navarro:2006mw}.

Furthermore we are restricting our study to the background evolution at low redshifts, 
assuming GR-like behaviour until then. We are not taking into account perturbations 
and their evolution, nor considering the stability of modes in the early universe, 
in constraining the models. These relevant issues have been studied in 
\cite{Sawicki:2007tf,Faraoni:2005vk,Tsujikawa:2007gd,Dolgov:2003px,Nunez:2004ts,Chiba:2005nz,DeFelice:2006pg,Calcagni:2006ye}.

This paper is organized as follows. In Section II we discuss the general metric based $f(R)$ case and its solutions. In 
Section III we discuss the Palatini formalism and its solutions. 
In Section IV we consider the generalization 
including the Gauss-Bonnet term. We end with a discussion and 
conclusions section. Appendices follow.

\section{General metric-based $f(R)$ model}

Let us begin with an $f (R)$ theory with the following action
\be
S = \int d^4 x \, \sqrt{- g} \, \frac{R + f
   (R)}{16 \pi G} + S_m\, , 
\label{STaction}
\ee
where $R$ is the Ricci scalar and $S_m$ is the action for the matter fields. The equations of motion are 
\[ (1 + f_R) G_{\mu \nu} - \frac{1}{2} g_{\mu \nu} (f - Rf_R) + g_{\mu \nu}
   \Box f_R - \nabla_{\mu} \nabla_{\nu} f_R = 8 \pi G\,T_{\mu \nu} ,\]
where an underscore $R$ implies a partial derivative with respect to $R$ ($f_R=\partial f/\partial R$). In a FRW background then,
\begin{equation}
  3 (1 + f_R) H^2 + \frac{1}{2} \,(f - R f_R ) + 3 H^2 f_R' = 8
  \pi G \rho\ , \label{fried0}
\end{equation}
where $H$ is the Hubble parameter and a prime denotes differentiation with respect to $N=\ln (a/a_0)$. Evaluated today this equation becomes
\begin{equation}
  1 + \beta+ \frac{1}{6} \, 
[\alpha - 6 \,\beta \, (H_0'/H_0 + 2)] + \gamma\, \frac{R_0'}{H_0^2} = \frac{8
  \pi G}{3H_0^2}\, \rho_0\ , \label{fried2}
\end{equation}
where an underscore 0 implies present values, and $\alpha=f_0/H_0^2$, $\beta=f_{R0}$, and $\gamma=f_{RR0}\, H_0^2$ are the dimension-less parameters of this theory. Since we have three parameters to solve for, we need three equations. These are obtained by differentiating the Friedmann equation twice. The equations then involve higher derivatives of $H$ evaluated today (in this case upto four).

Assuming that $f$ can be Taylor-expanded in $R$ about today and retaining only up to the second order terms (this being the simplest non-trivial case),
\begin{equation}
  f \approx f_0 + f_{R 0} \hspace{0.25em} H_0^2 \left[ \frac{R -
  R_0}{H_0^2}\right] + \frac{1}{2} \hspace{0.25em} f_{RR 0} \hspace{0.25em} H_0^4
  \left[ \frac{R - R_0}{H_0^2} \right]^{\! 2}\, .
\label{fRTaylor}
\end{equation}
In order to use the equations of motion effectively one needs to truncate the Taylor expansion at some order, otherwise one would need an infinite number of equations, obtained from repeatedly differentiating the Friedmann equation, to specify an infinite number of parameters, the Taylor coefficients. We chose to truncate this expansion of $f(R)$ at second order because stability constraints on these theories involve $f_R$ and $f_{RR}$ terms (see e.g.\ \cite{Sawicki:2007tf}).

Differentiating the Friedmann equation once gives
\be
6\,H\,H'\,(1+f_R)+3\,(H\,H'-H^2)\,f_R'+3\,H^2\,f_R''=8\pi G\,\rho'\, .
\label{Dfr1b}
\ee
 This relation together with the conservation of stress-energy (which ignoring contribution from radiation leads to)
\be
\rho'=-3\,(1+w)\,\rho\approx-3\,\rho\ .
\label{fRenergy}
\ee
is equivalent to the second Einstein equation. Equations (\ref{fRTaylor}), (\ref{Dfr1b}) and (\ref{fRenergy}) lead to
\be
6\,\tmH'\,(1+f_{R0})+3\,(\tmH'-1)\,f_{RR0}\,R'_0+3\,f_{RR0}\,R_0''=-9\,\Omega_m\, ,
\label{fRE2}
\ee
where we have defined the (present day) derivatives of the (normalized) Hubble parameter as 
\be
\tmH' = \frac{H'_0}{H_0}\,\qquad{\rm and}\qquad \tmH''=\frac{H_0''}{H_0}\, .
\ee

For a flat Friedmann-Robertson-Walker background  
\begin{eqnarray}
  R & = & 6\, (HH' + 2 H^2)\notag\\
R'&=&6\,({H'}^2+H\,H''+4\,H\,H')\label{DerivR}\\
R''&=&6\,(H\,H'''+3\,H'\,H''+4\,{H'}^2+4\,H\,H'')\notag\ ,
\end{eqnarray}
hence equation (\ref{fRE2}) can also be written as 
\be
\beta=
-\frac{2\tmH'+3\Omega_{m}+6\,\gamma\,
(\tmH'''+3\tmH''+4\tmH'\,\tmH''+{\tmH'}^3+7{\tmH'}^2-4\tmH')}{2\,\tmH'}
\label{betaG}
\ee

Differentiating the Friedmann equation a second time, evaluating it today and using equation (\ref{betaG}) gives
\bea
\gamma&=&\tfrac12\,\Omega_{m}\,(3\tmH'+{\tmH'}^2+\tmH'')\times
[9{\tmH'}^4+15{\tmH'}^2\tmH''+6{\tmH'}^3\tmH''\notag\\
&&\quad{}-3{\tmH''}^2
+3\tmH'\tmH'''+6{\tmH'}^2\tmH'''-\tmH''\tmH'''+\tmH'\tmH'''']^{-1}\ ,
\label{gammaG}
\eea
where
\be
\tmH'''\equiv\frac{H_0'''}{H_0}\qquad
{\rm and}\qquad
\tmH''''\equiv\frac{H_0''''}{H_0}\ .
\ee

We see that
\be
\alpha=\alpha(\tmH', \tmH'', \tmH''', \tmH'''', \Omega_{m})\ ,
\ee
and the same is true for $\beta$ and $\gamma$. The relations are non-linear. 

Posterior distributions of $\tmH'$, $\tmH''$, $\tmH'''$, $\tmH''''$ 
and $\Omega_{m}$ are obtained from a likelihood 
analysis of data in the following way. We use 182 Type Ia supernovae 
(SNe Ia) from the HST/GOODS program, together with first year SNLS and some nearby 
SNe Ia, as compiled by \cite{Riess07}. The $(R,l_a,\Omega_bh^2)$ 
combination, where $R$ and $l_a$ are CMB shift parameters\cite{Wang:2007mz},
is used to account for relevant constraints from the CMB 
\cite{Spergel:2006hy}. The SDSS baryon acoustic 
oscillation (BAO) scale measurement is also used \cite{E05}.  
Following the analysis method of \cite{Wang:2003gz}-\cite{Wang:2005ya},\cite{Wang:2007mz}, model independent constraints are derived on the Hubble parameter 
in linear redshift bins using a Markov Chain Monte Carlo (MCMC) 
algorithm. The $H(z)$'s of the MCMC chain elements are then 
converted into the derivatives of $\tmH$, as described in Appendix A. 
The top panel of Fig 1 shows the constraints thus derived 
on what we for convenience shall sometimes refer to as the ``initial parameters''.  
From the MCMC chains now we can obtain constraints on $\alpha$, $\beta$, and $\gamma$
using the equations derived above.

\begin{figure}[ht]
\begin{center}
{\psfrag{Hprime(1)}{${}_{\tmH'}$}
\psfrag{Hprime(2)}{${}_{\tmH''}$}
\psfrag{Hprime(3)}{${}_{\tmH'''}$}
\psfrag{Hprime(4)}{${}_{\tmH''''}$}
\psfrag{A}{${}_{\alpha}$}
\psfrag{B}{${}_{\beta}$}
\psfrag{C}{${}_{\gamma}$}
\includegraphics[width=12truecm]{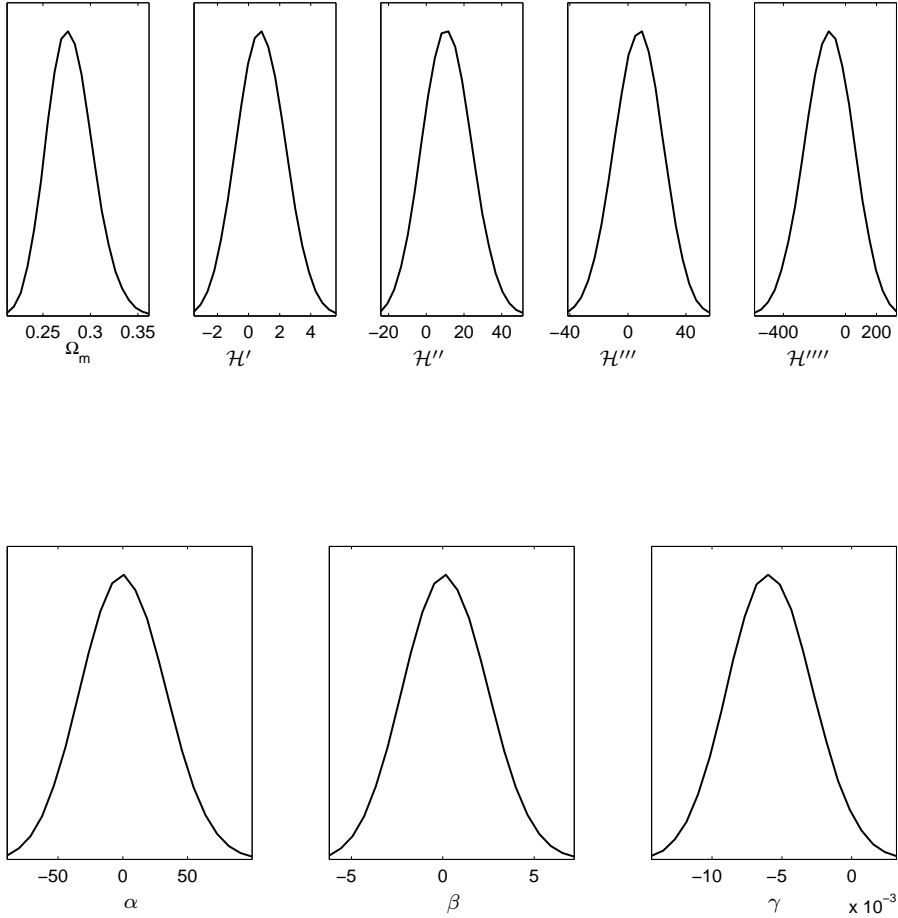}}
\label{fig:1}
\caption{Results for general metric-based $f(R)$ gravity: constraints on $\alpha=f_{0}/H_0^2$, $\beta=f_{R0}$, and $\gamma=f_{RR0}\,H_0^2$ (bottom panel), obtained from constraints on $\Omega_m$ and the derivatives of $\tmH$ (top panel), using current cosmological data.}
\end{center}
\end{figure}

\begin{figure}[ht]
\begin{center}
{\psfrag{tmH1}{$\tmH'''$}
\psfrag{tmH2}{$\;\,\,\,\,\,\,\tmH''''$}
%\psfrag{tmH2}[][][1][-90]{$\tmH''$}
\psfrag{Beta}{$\beta$}
\includegraphics[width=8truecm]{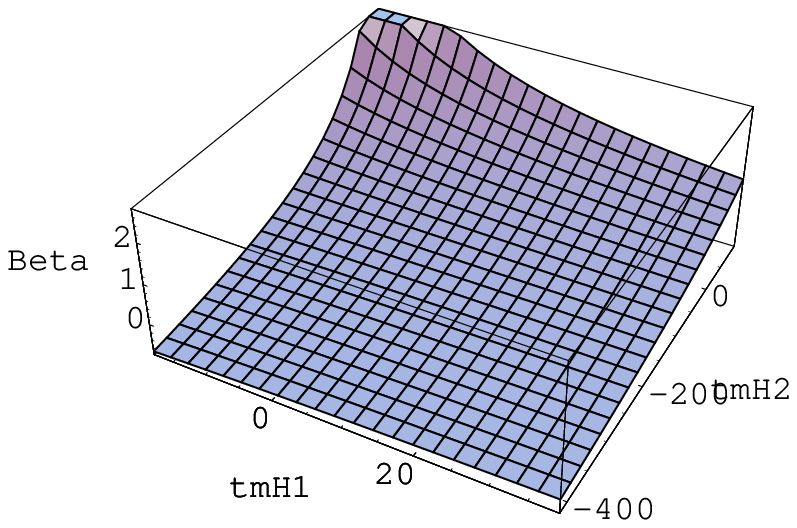}}
\label{fig2}
\caption{$\beta$ as a function of $\tmH'''$ and $\tmH''''$ setting $\tmH'$ and $\tmH''$ to their best values.}
\end{center}
\end{figure}

\subsubsection*{Linearization}

Equation (\ref{gammaG}) is highly non-linear in the initial parameters. 
Small changes in these parameters can have a large effect on the values 
of the $f(R)$ parameters, so that a relatively small number of high 
likelihood combinations of the initial parameters can lead to a large range 
of values for the $f(R)$ parameters. This leads to flat 1D parameter 
likelihood curves. This effect is suppressed in the MCMC posterior of
 the $f(R)$ parameters, because in addition to the likelihood this 
takes into account the number of samples that 
fall into each bin. In other words the discrepancy between these 
curves is due to there being a main posterior peak (representing 
a concentration of high likelihood points) together with isolated 
points in parameter space that are allowed by the 
likelihood. In addition, some allowed values of the initial parameters lead to 
singularities; these are the zero values of the $\tmH'$'s and the zeroes of 
denominator in the expression for $\gamma$. Given the discrete samples in the 
MCMC chain, the parameters $\alpha$, $\beta$ and $\gamma$ don't actually 
blow up, but instead as a result have a larger range. These problems should be 
at least somewhat eased by better data. For now, in order to avoid the 
discrepancy 
between the likelihood and the posterior, and to discount the region around 
singular points, we proceed to find solutions by first linearizing the 
equations for the modified gravity parameters about the mean values of 
the initial parameters, supported also by the fact that the initial 
parameters have close to Gaussian distributions. This approach will be 
used in subsequent sections as well.

We linearize the equations for $\alpha$, $\beta$ and $\gamma$ about their 
maximum likelihood (in MCMC, the relevant quantity is the mean) values, 
which are obtained in turn from the mean values of the initial parameters:
\bea
\alpha&\approx&\hat\alpha+(p_i-\hat p_i)\left.\frac{\partial\alpha}{\partial p_i}\right\vert_{p_j=\hat p_j}\, ,\label{linal}\\
\beta&\approx&\hat\beta+(p_i-\hat p_i)\left.\frac{\partial\beta}{\partial p_i}\right\vert_{p_j=\hat p_j}\, ,\\
\gamma&\approx&\hat\gamma+(p_i-\hat p_i)\left.\frac{\partial\gamma}{\partial p_i}\right\vert_{p_j=\hat p_j}\, ,\label{lingam}
\eea
where $p_i=(\tmH', \tmH'',\tmH''', \tmH'''', \Omega_{m})$ and hats represent 
mean values.

We obtain
\bea
\alpha&=&-29.7708 + 31.9923 \,\tmH' - 0.652908 \,\tmH'' - 0.0628292 \,\tmH'''\notag\\
&&\ {}+ 0.0513295 \,\tmH'''' + 63.1444 \,\Omega_{m}\ ,\\
\beta&=&-1.89148 + 2.29135 \,\tmH' - 0.0479787 \,\tmH'' - 0.00179758 \,\tmH'''\notag\\ \label{betaeqn} 
&&\ {}+ 0.00402273\,\tmH'''' + 4.20931 \,\Omega_{m}\ ,\\
\gamma&=&-0.00179067 - 0.00887807 \,\tmH' + 0.000625996 \,\tmH'' + 0.0000797733 \,\tmH'''\notag\\
&&\ {}  - 0.0000138531\,\tmH'''' - 0.02083 \,\Omega_{m}\, .
\eea

Even though the linearization is about the mean values of the initial 
parameters, we then use the MCMC chain to obtain the distributions of 
the modified gravity parameters. These are shown in the bottom panel of 
Fig 1. As described above these
represent the main body of solutions of the $f(R)$ theory allowed by the data; 
given the non-linearity of the equations there are other solutions isolated 
in parameter space. 
We see that while the order of magnitude of $\alpha$ and $\beta$ is constrained,
$\gamma\propto f_{RR,0}$ is found to be slightly negative today over most of 
its allowed
 range. Even though we are not discussing stability issues here, 
\cite{Sawicki:2007tf,Bean:2006up,Amendola:2006we,Song:2006ej} find that 
$f_{RR}$ at high redshift needs to be positive in order to avoid instability 
and obey GR. Therefore under this theory $\gamma$ would have to change sign 
at some $R$ before today. This is interesting.

Fig 2 shows the 
solutions for $\beta$ at the mean values of the better constrained 
derivatives $\tmH'$ and $\tmH''$, and over the allowed ranges of the 
higher derivatives. A range of values for $\beta$ are possible, 
including the special case of $\beta=1/3$ discussed below.

\subsection{Metric based $f(R)$ with $\beta=1/3$}

If we assume that in the solar system the real metric can be expanded about 
GR Schwarzchild, then for the previous action, the effective Newton's constant 
can be written as \cite{Torres:2002pe}
\[
G_{\text{eff}} = \frac{4}{3} \, \frac{G}{1 + f_R} \,,
\]
which can be recast as a constraint on $\beta$ today (imposing $G=G_{\rm eff,0}$)
\begin{equation}
  f_{R0} = \beta=\tfrac{1}{3}\  .\label{G0}
\end{equation}

Imposing $\beta=1/3$, equation (\ref{fried2}) and (\ref{betaG}) give
\begin{equation}
  \tfrac{1}{6} \, \alpha - \tfrac{1}{3} (\mathcal{H}' + 2) + 6 \hspace{0.25em}
  \gamma \hspace{0.25em} ( \tmH'' + {\tmH'}^2 + 4
  \mathcal{H}') = \Omega_{m} - \tfrac{4}{3}\, , \label{eq:param1}
\end{equation}
and
\begin{equation}
  \tfrac{8}{3} \, \mathcal{H}' + 6 \gamma \left[ 3
  \tmH'' + \tmH' \left( {\tmH'}^2 + 7 \tmH' + 4 \tmH'' - 4 \right) +
  \tmH''' \right] = - 3 \hspace{0.25em} \Omega_{m} \label{eq:param2} .
\end{equation}
The data define constraints on the $\tmH'$'s as discussed in the previous section. Equations~(\ref{eq:param1})
and~(\ref{eq:param2}) can then be used to solve for $\alpha$ and $\gamma$. The complications arising due to the equations being non-linear in the initial parameters, as discussed earlier, apply here as well. Hence as before we proceed to linearize the equations of motion about the mean values of the $\tmH'$'s and $\Omega_m$.

\subsubsection*{Linearization}

Using equations (\ref{linal}) and (\ref{lingam}) together with equations (\ref{eq:param1}) and (\ref{eq:param2}), we get
\bea
\alpha&\approx&-4.53219 + 12.1526 \,\tmH' - 1.54509 \,\tmH'' - 0.974158 \,\tmH''' + 20.5967 \Omega_m\, ,\\
\gamma&\approx&-0.028308 - 0.0521268 \,\tmH' + 0.0262386 \,\tmH'' + 0.00930701 \,\tmH''' - 0.139456 \Omega_m\, .
\eea
These relations, used on the MCMC chains, give the results shown in Fig 3. 
\begin{figure}[ht]
\begin{center}
{\psfrag{Hprime(1)}{${}_{\tmH'}$}
\psfrag{Hprime(2)}{${}_{\tmH''}$}
\psfrag{Hprime(3)}{${}_{\tmH'''}$}
\psfrag{Hprime(4)}{${}_{\tmH''''}$}
\psfrag{A}{${}_{\alpha}$}
\psfrag{B}{${}_{\beta}$}
\psfrag{C}{${}_{\gamma}$}
\includegraphics[width=12truecm]{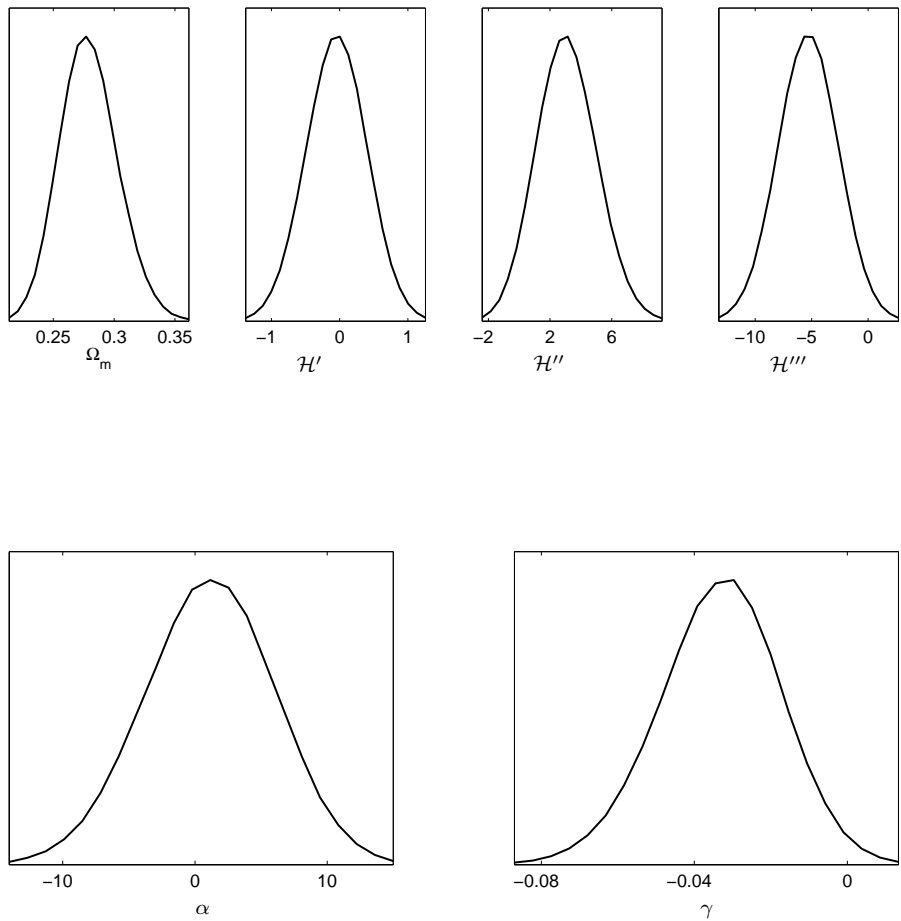}}
\label{fig3}
\caption{Same as Fig 1 for the $\beta=\tfrac13$ case of the general metric-based $f(R)$ theories.}
\end{center}
\end{figure}

\section{$f(R)$ in the Palatini formalism}

So far we have considered theories for which the gravity variables were chosen to be the metric elements $g_{\mu\nu}$. On the other hand, after writing down the action, one is free to choose a different set of fields. For example, one can choose to find the equations of motion by varying the action with respect to the following two tensorial quantities, $\delta g_{\mu\nu}$ and $\delta\Gamma^\lambda_{\mu\nu}$, which are the metric and the Christoffel symbols perturbations respectively. In standard GR, the approach of choosing two different field variables, leads to the same standard Einstein equations of motion. However, in $f(R)$ theories, with $f_R\neq {\rm constant}$, the equations of motion are indeed different. Of course, one may wonder which description of gravity is the correct one. However, since gravity is the least well known force, many physicists have argued that the possibility of introducing extra fields should be considered.

In this formalism (refered to as Palatini, also studied in \cite{Fay:2007gg,Sotiriou:2005hu,Olmo:2004hj,Capozziello:2004vh}), $R_{\mu\nu}$ becomes a function only of the Christoffel symbols, so that it cannot be written as usual in terms of derivatives of the scale factor. In this case it is customary to write the general action in the following way
\be
S=\int d^4x\,\sqrt{-g}\,\frac{f(R)}{16\pi G}+S_m\, .
\ee
The Einstein equations obtained by varying the action with respect to the metric elements are
\be
f_R\, R_{\mu\nu}-\tfrac12\,f\,g_{\mu\nu}=8\pi G\,T_{\mu\nu}\, ,
\label{Pala00}
\ee
the trace of which is
\be
f_R\,R-2\, f=-8\pi G\,\rho_m\,.
\label{Pala0}
\ee
Taylor-expanding $f$ in terms of $R$ about today, as before, we have
\be
\bar f\equiv\frac{f}{H_0^2}=\alpha+\beta\,(\bar R-\bar R_0)+\tfrac12\,\gamma\,(\bar R-\bar R_0)^2\, ,
\ee
where $\bar R\equiv R/H_0^2$, and $\bar R_0$ is its value today.
Therefore from equation (\ref{Pala0}) 
\be
\bar R_0=\frac{2\,\alpha-3\,\Omega_m}\beta
\ee
and in general
\be
\bar R\equiv\frac{R}{H_0^2}=\frac{2\alpha-2\beta\bar R_0+\gamma\,\bar R_0^2-3\Omega_m\,e^{-3N}}%
{\gamma\,\bar R_0-\beta}\, ,
\ee
This relation is important as we know how $R$ varies close to today. 
Taking the derivative of $R$ with repect to time, plugging it into the 00 component of equation (\ref{Pala00}), we get the new Friedmann equation
\be
\frac{H^2}{H_0^2}=\frac{6\,\Omega_m\,e^{-3N}+f_R\,\bar R-\bar f}{6\,f_R\,\xi}\, ,
\label{PalaFR}
\ee
with $H=H(N)$ is the Hubble factor $H=a^{-1}da/dt$, and
\be
\xi=\frac{[2\,e^{3N}\,(\beta^2-2\alpha\,\gamma)-3\gamma\,\Omega_m]^2}%
{4\,[e^{3N}\,(\beta^2-2\alpha\,\gamma)+3\gamma\,\Omega_m]^2}\, 
\ee
following \cite{Fay:2007gg}.

The Friedmann equation (\ref{PalaFR}) together with its two
$N$-derivatives give us three equations for three unknowns. 
In contrast to the metric-based case, the Friedmann equation here does not involve any derivatives of the Hubble parameter, because $R$ is independent of $H$. 
We end up with three non-linear equations involving $\Omega_m$, $\tmH'$, $\tmH''$, to be solved simultaneously 
for $\alpha$, $\beta$ and $\gamma$. This time though, the equations are non-linear even in $\alpha$, $\beta$, and $\gamma$, 
so that there isn't a unique solution for these parameters.
In general there is no telling how many real 
solutions there are. We solve the equations numerically 
in the range [-10,10] for the parameters.  Listed in 
Appendix B are multiple solutions found at the mean values of the initial parameters. Next we consider one of these solutions.

\subsection*{Linearization}

The equations of motion can be written as $F_i(p_j, q_k)=0$, with $i=1,2,3$, where $F_1$=0 corresponds to the Friedmann equation, and $F_2$=0, and $F_3$=0 to its two derivatives. Let $p_j=(\alpha, \beta, \gamma)$ and $q_k=(\Omega_m, \tmH', \tmH'')$. The solutions listed in Appendix B are the $\hat p_j$ which solve $F_i(\hat p_j,\hat q_k)=0$, where $\hat q_k$ correspond to the mean values of the initial parameters. About any one solution we may linearize the equations of motion (for reasons explained in section II):
\be
F_i(p_j,q_k)=0\approx F_i(\hat p_j,\hat q_k)+\left.\frac{\partial F_i}{\partial p_j}\right\vert_{\hat p,\hat q}(p_j-\hat p_j)
+\left.\frac{\partial F_i}{\partial q_j}\right\vert_{\hat p,\hat q}(q_j-\hat q_j)\, .
\ee
With $A_{ij}=\partial F_i/\partial p_j\vert_{\hat p,\hat q}$ and $B_{ik}=\partial F_i/\partial q_k\vert_{\hat p,\hat q}$, one has
\be
p_j=-(A^{-1}\,B)_{jk}\,q_k+\hat p_j+(A^{-1}\,B)_{jk}\,\hat q_k\, .
\label{mateqn}
\ee 

As an example, consider the solution [0.222102, 0.00488155, 4.12207e-05]. We can obtain the distributions of $p_j$ about this solution using the linearized equations together on the MCMC chains for the initial parameters. Results are shown in Fig 4, and for this solution the matrix $A^{-1}\,B$ is also given in Appendix B. 
Though the figure shows reasonably strong constraints on the modified gravity parameters, it should be remembered that in this formalism the data allow multiple such solutions. One may try to distinguish between the solutions by imposing consistency checks derived from taking higher derivatives of the Friedmann equation. In that case we would use constraints on the Hubble parameter in more redshift bins, which would allow more freedom and hence more solutions overall to distinguish between (even though here we are only discussing solutions about one set of values for the initial parameters). Hence this endevour will not be fruitful at this time. With much better data, the solutions may be tractable. 

\begin{figure}[ht]
\begin{center}
{\psfrag{Hprime(1)}{${}_{\tmH'}$}
\psfrag{Hprime(2)}{${}_{\tmH''}$}
\psfrag{Hprime(3)}{${}_{\tmH'''}$}
\psfrag{Hprime(4)}{${}_{\tmH''''}$}
\psfrag{A}{${}_{\alpha}$}
\psfrag{B}{${}_{\beta}$}
\psfrag{C}{${}_{\gamma}$}
\includegraphics[width=10truecm]{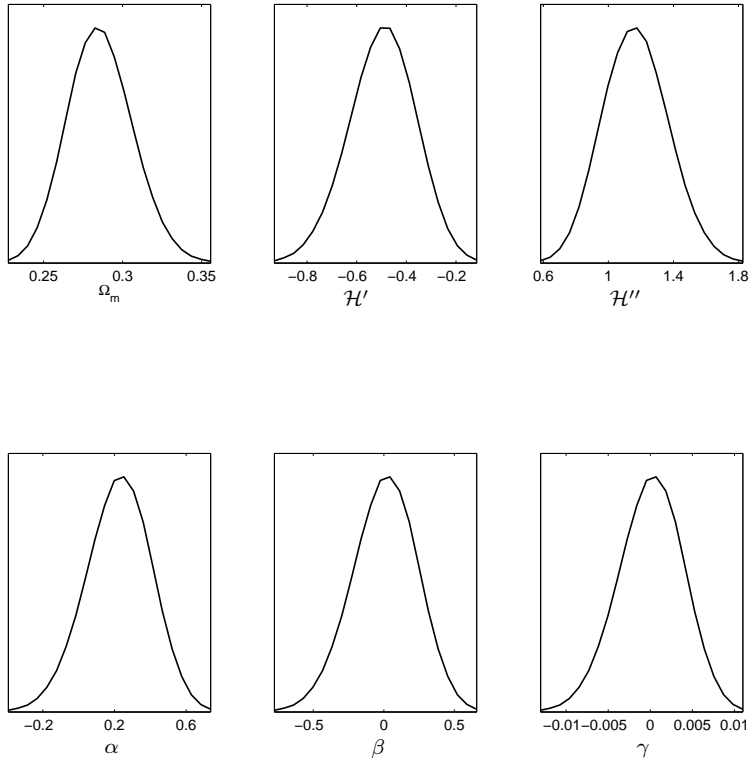}}
\label{fig:Pal}
\caption{Same as Fig 1, for $f(R)$ gravity in the Palatini formalism. The constraints in the bottom panel are about one solution of the three non-linear equations of motion. There are other solutions, see text for further discussion.}
\end{center}
\end{figure}

We also note in passing that when we constrain just two $\tmH'$'s, these turn out to be constrained to non-zero values (Fig 4).  A negative $\tmH'$ implies that the universe is not super-accelerating today, and positive $\tmH''$ implies that $\tmH'$ is increasing (approaching zero) today. This is as expected. When more $\tmH'$'s are involved, as in the remaining sections, they of course become less constrained.
  
% put plot

\section{$f(R, \GB)$ metric-based theories}

It would be interesting to look also at other generalizations which have been considered in the literature \cite{Carroll:2004de}-\cite{DeFelice:2006pg}. These models have actions of the form
\be
S=\frac1{16\pi G}\int d^4x\,\sqrt{-g}\,[R+ f(R, \GB)]+S_m\, ,
\ee
where $\GB=R^2-4\,R_{\mu\nu}\,R^{\mu\nu}+R_{\mu\nu\rho\sigma}\,R^{\mu\nu\rho\sigma}$ is the Gauss-Bonnet combination. The equations of motion which follow can be written as
\bea
&&(1+F)\,G_{\a\b}-\N_\a\N_\b F +g_{\a\b}\,\Box F -\tfrac12\, g_{\a\b}\,(f-F\,R-\xi\,\GB)
\notag\\
&&\quad{}-2\,R\,\N_\a\N_\b\xi +2\,R\,g_{\a\b}\Box\xi
-4\,R_{\a\b}\,\Box\xi-4\,R_{(\a}{}^{\s\tau}{}_{\b)}\,\N_\s\N_\tau\xi\notag\\
&&\quad{}-4\,g_{\a\b}\,R^{\rho\s}\,\N_\rho\N_\s\xi+8\,R_{(\a}{}^\nu\N_{\b)}\N_\nu\xi=8\pi G\,T_{\a\b}\, ,
\eea
% These equations can be found by using the general equation in De Felice, Trodden, Hindmarsh
% where U = -b\s^2 f +\l (1+F) +\phi \xi -l-f, 2b\s f=F, \g=1.
where $F=\partial f/\partial R$ and $\xi=\partial f/\partial \GB$ and the partial derivatives should be found treating $R$ and $\GB$ as independent variables. In a FRW background this becomes
\be
3\,H^2\,(1+F+F') +\tfrac12\,(f-F\,R-\xi\,\GB)+12\,H^4\,\xi'=8\pi G\,\rho\, ,
\label{eq:GSC}
\ee
where
\be
\GB=24\,\frac{\ddot a}a\,H^2=24\,H^3\,(H'+H)\, ,
\ee
and $R$ has been already introduced in equations~(\ref{DerivR}).

The Taylor-expansion of $f$ about today up to second order in the scalars takes the form
\bea
\frac f{H_0^2} &=& \alpha+\beta\,(\bar R-\bar R_0)
+\tfrac12\,\gamma_1\,(\bar R-\bar R_0)^2\notag\\
&&\quad{}+\gamma_2\,(\bar R-\bar R_0)(\bGB-\bGBo)
+\tfrac12\,\gamma_3\,(\bGB-\bGBo)^2\, ,
\eea
where
\be
\alpha\equiv\frac{f_0}{H_0^2}\, , \;
\beta\equiv F_0\, , \;
\gamma_1\equiv H_0^2\left.\frac{\partial F}{\partial R}\right\vert_0\, , \;
\gamma_2\equiv H_0^4\left.\frac{\partial\xi}{\partial R}\right\vert_0\, , \;
\gamma_3\equiv H_0^6\left.\frac{\partial\xi}{\partial\GB}\right\vert_0\, ,
\ee
with $\bar R=R/H_0^2$ and $\bGB = \GB/H_0^4$. It should be noted that there is no linear term in the expansion for $\GB$, because such a term would give no contribution to the equations of motion, and the constant term in $\bGB\vert_0$ is considered absorbed in $\alpha$. Then one has
\bea
F&=&\beta+\gamma_1\,(\bar R-\bar R_0)+\gamma_2\,(\bGB-\bGBo)\, ,\\
H_0^2\,\xi&=&\gamma_2\,(\bar R-\bar R_0)+\gamma_3\,(\bGB-\bGBo)\, .
\eea

Today equation (\ref{eq:GSC}) becomes
\bea
\alpha&=&6\,[-1 + (1+\tmH')\, \beta  -6\gamma_1\,(4\,\tmH' + {\tmH'}^2 +  
        \tmH'')\notag\\
&&\quad{} -48\,\gamma_2\,( 4\,\tmH' + 2\,{\tmH'}^2+ 
        \tmH'')\notag\\
&&\quad{} -96\,\gamma_3\,( 4\,\tmH' + 3\,{\tmH'}^2 
         + \tmH'') + \Omega_m]\, .
\eea
Taking four derivatives of equation (\ref{eq:GSC}) allows us to solve for the five parameters in term of $\Omega_m$ and the the derivatives of $\tmH$ (up to the 6th derivative). One derivative gives
\bea
\beta&=& -1 + (2\tmH')^{-1}[6(4\tmH' - 7\tmH'^2 - \tmH'^3 - 3\tmH'' - 4\tmH'\tmH'' - \tmH''')\,\gamma_1] \notag\\
&&\quad{}+ (2\tmH')^{-1}[48(4\tmH' - 14\tmH'^2 - 6\tmH'^3 - 3\tmH'' - 8\tmH'\tmH''  - \tmH''' )\gamma_2]\notag \\
&&\quad{}+(2\tmH')^{-1}[96(4\tmH' - 21\tmH'^2 - 15\tmH'^3 - 3\tmH'' - 12\tmH'\tmH''  - \tmH''' )\gamma_3]-\frac{3\Omega_m}{2\tmH'} 
\eea
Further derivatives give $\gamma_1$, $\gamma_2$ and $\gamma_3$. The later two particularly involve extremely complicated expressions that cannot be written down here.

When constraining 6 derivatives of the Hubble parameter from data, it is expected that the constraints will be poor. The top two panels of Fig 5 show the constraints on the initial parameters. For reasons discussed in section II, we will now linearize the relations for the modified gravity parameters about the mean likelihood values of the inital parameters. In this way we illustrate a space of possible solutions. (There would be a number of isolated solutions not represented in these results, but because these are isolated solutions they will also be suppressed in the full posterior. Better data should reduce the number of such solutions.) 

\begin{figure}[ht]
\begin{center}
{\psfrag{Hprime(1)}{${}_{\tmH'}$}
\psfrag{Hprime(2)}{${}_{\tmH''}$}
\psfrag{Hprime(3)}{${}_{\tmH'''}$}

\psfrag{Hprime(4)}{${}_{\tmH''''}$}
\psfrag{Hprime(5)}{${}_{\tmH'''''}$}
\psfrag{Hprime(6)}{${}_{\tmH''''''}$}
\psfrag{A}{${}_{\alpha}$}
\psfrag{B}{${}_{\beta}$}
\psfrag{C1}{${}_{\gamma_1}$}
\psfrag{C2}{${}_{\gamma_2}$}
\psfrag{C3}{${}_{\gamma_3}$}
\includegraphics[width=10truecm]{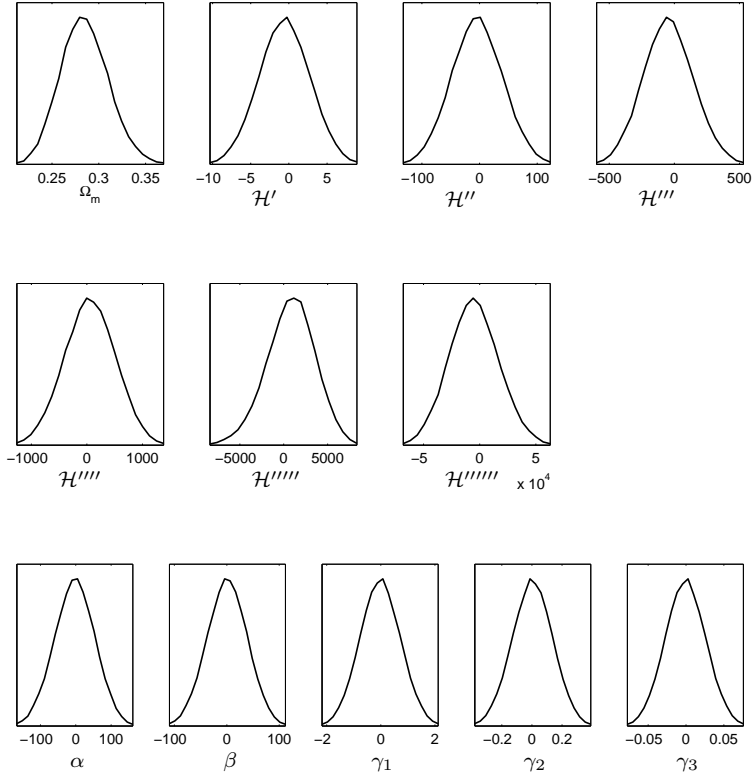}}
\label{fig:fgb}
\caption{Same as Fig 1, for $f(R,\GB)$ gravity in the metric formalism.}
\end{center}
\end{figure}

\subsection*{Linearization}

Linearizing the previous equations about the mean values of the the derivatives of $\tmH$ and $\Omega_m$ one finds
\bea
\alpha& =& -4.52918 + 14.6151 \tmH' - 3.28575 \tmH'' + -0.200232 \tmH'''\notag\\
&&\quad{} - 0.006872 \tmH'''' + 0.0004706 \tmH''''' \
+ 8.5558\times10^{-6} \tmH'''''' + 18.3942 \Omega_m \\
\beta&=& -0.42128 + 4.91222 \tmH' - 1.71918 \tmH'' + 0.1071
\tmH''' \notag\\
&&\quad{}- 0.003735 \tmH'''' + 0.000257 \tmH''''' + 4.6535\times10^{-6} \tmH'''''' + 5.39085 \Omega_m
\eea

\bea
\gamma_1&=& -0.008276 -
0.05569 \tmH' + 0.03140 \tmH''-
0.002392 \tmH''' \notag\\
&&\quad{} + 0.00004811 \tmH'''' -
6.2764\times10^{-6} \tmH''''' - 1.03197\times10^{-7} \tmH'''''' - 0.0705 \Omega_m\\
\gamma_2&=& 0.0006553 + 0.007342 \tmH' - 0.005405 \tmH''+ 0.000428 \tmH'''-
6.2403\times10^{-6} \tmH''''\notag\\
&&\quad{}   + 1.2073\times10^{-6} \tmH''''' + 1.9076\times10^{-8} \tmH'''''' + 0.01237 \Omega_m\\
\gamma_3&=& 0.0003048 -
0.000411 \tmH' + 0.0009694 \tmH'' - 0.00008026 \tmH'''+ 6.99103\times10^{-7} \tmH''''\notag\\
&&\quad{}  - 2.5368\times10^{-7} \tmH''''' 
- 3.8354\times10^{-9} \tmH'''''' - 0.001939 \Omega_m
\eea
These relations together with the MCMC chains result in the constraints shown in the bottom panel of Fig 5. The parameters are all consistent with zero, the significance is in the order of the constraints.

We have checked using the full equations that the $\gamma$'s are in fact well constrained to be close to zero as indicated by the plot. This is because the relations contain combinations of the derivatives of $\tmH$ in their denominators. Hence this is a consequence of the equations of motion. This result is interesting. Again the $\tmH'$'s=0 is a singular point for the $\gamma$'s (as well as for $\alpha$ and $\beta$ the relations for which contain a division by $\tmH'$). Again there is a solution allowed by data that is a singular point in this theory, but overall (ie. for most allowed combinations of the derivatives of $H$) the parameters of the theory are well behaved, and the distributions shown encompass the main range of values for these parameters that are acceptable as per current data.

\section{Discussion and conclusions}

$f(R)$ and $f(R,\GB)$ models, by definition, introduce higher
derivatives of the expansion rate into the equations of motion. This
automatically requires data which can allow for precise measurements
of derivatives of the Hubble parameter. This severe demand on
observations arises because currently we have no theory and no
symmetry which can exactly predict the form of $f(R)$. This problem is
evidently similar to the task of finding the form for the potential of
a quintessence-like field, or more generally to finding a dynamical
alternative to a cosmological constant. On the other hand, we should
not feel so complacent as to accept a tiny cosmological constant as
the solution to the dark energy problem without giving due consideration to the
daunting dynamical alternatives in the gravity sector.  In order to get 
rid of the
possibilities discussed in this paper of modifying gravity, one may
introduce the axiom that only a linear combination of Lovelock terms
can enter into the Lagrangian for gravity, although this axiom does
not follow from any symmetry consideration. However, this would still
not be enough to solve the cosmological constant problem, namely the
zeroth Lovelock term, the cosmological constant that we would predict
from QFT is far too large compared to the one needed by
observations. $f(R)$ models, or more generally $f(R,\GB)$ theories,
thus remain a relevant consideration.

We have used distance measurements from CMB, BAO and SNe Ia to place
preliminary bounds on the parameters of $f(R)$ models. $f(R)$
was Taylor expanded about today, keeping terms upto second order in
$R$. Equations of motion were found in both the metric and Palatini 
formalisms. These allow us to solve for the coefficients of the 
Taylor expansion of $f(R)$. The equations of motion are non-linear, 
and we study the main space of possible solutions. The solutions are 
interesting in that they reflect the order of magnitude of the 
coefficients allowed by data. Under the metric formalism we
find that $\partial f/\partial R\vert_0$ is small and negative 
over a substantial part of the allowed region, in which case in
order for these models to have a consistent GR-like evolution at 
early times(see \cite{Sawicki:2007tf}), this quantity must have changed 
sign some time in the past.

We have also analyzed a more general class of theories, $f(R,
\GB)$. In order to modify gravity, without introducing spurious
degrees of freedom (such as extra ghost-like spin-2 particles), one
needs to use Lovelock scalars \cite{Hindawi:1995an, Nunez:2004ts}. In
four dimensions the only such terms which give a non-zero contribution
are a constant, $R$, and $\GB$. These models are more general and the 
constraints are weaker. However, as for the $f(R)$
theories, we should make a serious attempt to shrink the allowed
parameter space for such theories in different ways (using data and 
theoretical considerations).

In addition to the analysis presented here, at the moment, other
considerations (convergence to GR-like evolution at early times,
instabilities) are still the most serious issues that these theories
need to survive.  For example, it is known that some models for $f(R)$
and $f(R,\GB)$ do not have a FRW background consistent with GR at
early times \cite{Amendola:2006we, DeFelice:2007zq}. Furthermore at
high-redshift GR-like models may lead to unstable behaviour in the
evolution of perturbations, because of the existence of either a
tachionic mode or a mode with imaginary speed of propagation
\cite{Sawicki:2007tf, DeFelice:2006pg, Li07}. Such considerations will
further constrain the parameter space of the models considered here.
In this paper we have also not considered solar system type
constraints, for reasons discussed in the Introduction.

To conclude, we have used current data to place constraints on the
first and second derivatives of general functions of the Ricci and the 
Gauss-Bonnet scalars, under different formalisms. These derivatives are
 important quantities for theoretical considerations relating to ghosts and 
instabilities. Next generation dark energy surveys \cite{detf,ground,jedi}
can measure the cosmic expansion history much more precisely; 
this would dramatically shrink the presently allowed parameter 
space of the modified gravity models considered here.

\section{Acknowledgements}
We thank Andrew Liddle and Mark Hindmarsh for helpful comments. ADF and PM are supported by STFC, UK.
YW is supported in part by NSF CAREER grants AST-0094335.

\appendix

\section{Inverting $H(z)$'s to the derivatives of $\tmH$}

Consider the vector $h_i=\tmH_i-1=H(z_i)/H_0-1$. If the Hubble parameter is measured in $n$ redshift bins, one can solve for $n$ of its (present day) derivatives by assuming a truncated Taylor expansion for each of $h_i$ about today upto order $n$, which gives
\begin{equation}
h_i\approx\sum_{j=1}^{n}\left.\frac{d^j\tmH}{dz^j}\right|_{z=0} \frac{z_i^j}{j!}\ .
\end{equation}
This is a linear equation, $h_i=A_{ij}\left.\frac{d^j\tmH}{dz^j}\right|_0$, where $A_{ij}=z^j_i/j!$, which can be inverted to find derivatives of $\tmH$ (today). We then change variables from $z$ to $N$, where $N = -\ln(1+z)$. One has
\begin{equation}
\frac{d^i\tmH}{dz^i}=(-1)^i\left[e^N\,\frac d{dN}\right]^{\!i}\tmH\,
\end{equation}
so that
\begin{eqnarray}
\left.\frac{d\tmH}{dz}\right|_0&=&-\tmH'_0\\
\left.\frac{d^2\tmH}{dz^2}\right|_0&=&\tmH''_0+\tmH'_0\\
\left.\frac{d^3\tmH}{dz^3}\right|_0&=&-(\tmH^{(3)}_0+3\,\tmH''_0+2\,\tmH'_0)\\
\left.\frac{d^4\tmH}{dz^4}\right|_0&=&\tmH^{(4)}_0+6\,\tmH^{(3)}_0+11\,\tmH''_0+6\,\tmH'_0\\
\left.\frac{d^5\tmH}{dz^5}\right|_0&=&-(\tmH^{(5)}_0+10\,\tmH^{(4)}_0+35\,\tmH^{(3)}_0+50\,\tmH''_0+24\,\tmH'_0)\\
\left.\frac{d^6\tmH}{dz^6}\right|_0&=&\tmH^{(6)}_0+15\,\tmH^{(5)}_0+85\,\tmH^{(4)}_0+225\,\tmH^{(3)}_0+274\,\tmH''_0+120\,\tmH'_0\ .
\end{eqnarray}

In fact it can be shown that 
\begin{equation}
\left.\frac{d^k\tmH}{dz^k}\right|_0=(-1)^k\,\sum_{j=1}^k a_{kj}\,\tmH^{(j )}_0\ ,
\end{equation}
with
\begin{eqnarray}
a_{k1}&=&(k-1)!\\
a_{kj}&=&(k-1)\,a_{k-1,j}+a_{k-1,j-1}\ ,\qquad 1<j<k\\
a_{kk}&=&1\ .
\end{eqnarray}
This then specifies the whole series of polynomials.

Writing the linear relation as
\begin{equation}
\left.\frac{d^j\tmH}{dz^j}\right|_0=B_{jk}\,\tmH^{(k)}_0\ ,
\end{equation}
we have
\begin{equation}
h_i=A_{ij}\,B_{jk}\,\tmH^{(k)}_0\equiv M^{-1}_{ik}\,\tmH^{(k)}_0\ ,
\end{equation}
or
\begin{equation}
\tmH^{(k)}_0=M_{kj}\,h_j\ .
\end{equation}

For two bins case ($z_1=0.7$, and $z_2=1.4$ for current data) we get
\be
M=\left( \begin {array}{cc} - 2.85714& 0.714286\\
\noalign{\medskip}- 1.22449& 1.32653
\end{array} \right) 
\ee
For three bins ($z_1=0.4667$, $z_2=0.9333$, and $z_3=1.4$) then we get
\be
M=\left( \begin {array}{ccc} - 6.4286& 3.2143
&- 0.71429\\\noalign{\medskip}- 16.531&
 15.153&- 3.8775\\\noalign{\medskip}
 32.930&- 22.369& 3.2216
\end {array} \right)
\ee
for 4 bins equally spaced out to $z=1.4$, we get 
\be
M=\left( \begin {array}{cccc} - 11.429& 8.5714&- 3.8095& 0.71428
\\\noalign{\medskip}- 59.320& 68.980&-34.286& 6.7687
\\\noalign{\medskip}- 9.0962& 55.802&- 52.789&13.251
\\\noalign{\medskip} 509.11&-745.18& 450.18&- 91.607
\end {array} \right)
\ee 
for 5 bins, we have
\be
M=\left( \begin {array}{ccccc} 
- 17.857& 17.857&- 11.905& 4.4647&- 0.71429
\\\noalign{\medskip}
- 145.83&209.61&- 153.91& 60.374&-9.9150
\\\noalign{\medskip}
- 335.37& 679.30&- 630.53& 276.88&-48.546
\\\noalign{\medskip} 
1445.8&- 2258.6& 1643.0&- 562.54& 79.241
\\\noalign{\medskip} 
2094.7&- 6287.7& 7809.3&- 4285.9& 838.55
\end {array} \right)
\ee

and, finally, for 6 bins, we have
\be
M= \left( \begin {array}{cccccc} - 19.3&
 16.109&- 7.193& 0.03758&
 1.271&- 0.3546\\\noalign{\medskip}-
 179.35& 218.78&- 108.04&
 0.67106& 20.4867&- 5.823
\\\noalign{\medskip}- 500.66& 833.8&-
 523.69& 4.9750& 118.16&-
 35.142\\\noalign{\medskip} 1837.10&-
 2403.5& 1168.7& 4.745&-
 149.15& 32.4457\\\noalign{\medskip}
 4187.9&- 9101.2& 7205.2&-
 112.35&- 2310.5& 732.79
\\\noalign{\medskip}- 54953& 91546&-
 59415& 191.52& 14897&-
 4190.4\end {array} \right)
\ee
With the present day derivatives of the Hubble parameter thus derived from model independent measurements of the Hubble parameter in linear redshift bins from cosmological data, we can then obtain constraints on parameters of modified gravity models, using appropriate equations of motion. This is done in sections II,III and IV of this paper.

\section{Appendix B: Solutions for the Palatini formalism}

We could find the following distinct solutions of the equations $F_i(\bar p_j,\bar q_k)=0$ for  $\alpha$, $\beta$, and $\gamma$, about the mean values for $\bar q_k=(\Omega_m. \tmH',\tmH'')$: [-0.857943, -0.428971, -0.14299],
[-1.46637, -0.422284, -0.140728],
[0.399627, 0.824305, 0.852729],
[3.78266, 0.748323, -0.00792803],
[0.222102, 0.00488155, 4.12207e-05],
[-0.739164, -0.00165984,  0.0000280094],
[-0.866737, -0.00527848,  0.000315044],
[0.252698,  -0.00306054,  0.0000152404],
[-0.29676,  0.00333869,  0.0000390608],
[-0.811148,  -0.00218394,  0.0000614933],
[-0.410838,  -0.0124848,  -0.0000849403],
[-0.132685,  -0.00276524,  0.0000205499], and
[0.00810481, -0.00504213,  -0.00002751].

For the solution [0.222102, 0.00488155, 4.12207e-05], the $A^{-1}\,B$ of equation (\ref{mateqn}) is 
\be
A^{-1}\,B=\left(\begin{array}{ccc}
          8.7943&-0.0077692& 0.000655508\\
          11.162&-0.015052&0.0012888\\
          0.18687&-0.00025239& 0.000021581
\end{array}\right).
\ee

\end{document}